\documentstyle[preprint,12pt,aps,prl]{revtex}
\begin{document}
\preprint{ISN--93--50}
\draft
\tighten
\title{ Proof of stability of the hydrogen molecule}
\author{J.-M. Richard}
\address{Institut des Sciences Nucl\'eaires,
Universit\'e Joseph Fourier - CNRS - IN2P3\\
53, avenue des Martyrs, Grenoble, France}
\author{J. Fr\"ohlich, G.-M. Graf, M. Seifert}
\address{Theoretical Physics, ETH, H\"oggerberg,
Z\"urich, Switzerland}
\date{\today}

\maketitle

\begin{abstract}
We sketch two rigorous proofs of the stability of the
hydrogen molecule

in quantum mechanics.
 The first one is based on an extrapolation of
variational estimates of

the groundstate energy of
a  positronium
molecule to arbitrary mass ratios. The second one is
an
extension of Heitler-London theory to nuclei of finite
mass.
\end{abstract}
\pacs{03.65.Ge, 31.15.+ q, 31.20. Di, 36.10. Dr, 12.40.
Qq.}

The stability of the hydrogen molecule H$_2

\left({ pp e^- e^- }\right)$ plays an important role in our
understanding of
chemical binding and thus deserves a mathematically
precise analysis. It
appears that, within the usual formalism of
non-relativistic quantum mechanics,
a rigorous proof of the fact that the system

$\left({ pp e^- e^- }\right)$ has bound states, i.e., that
H$_2$ is stable, is
not available in the literature, at least to our
knowledge.

In this letter, we sketch two somewhat complementary
proofs of the stability of
H$_2$. The first proof starts from a
four-body system consisting of two particles of charge
+ 1 and two particles of
charge $- 1$, all of which have the same mass, i.e.,
from a system corresponding
to a positronium molecule, the stability of which has
been essentially
established by Hylleraas and Ore \cite{Ore}, up to a
subtlety concerning the threshold
of the continuous spectrum.

An elementary variational argument can then be used
to extrapolate upper bounds

on the groundstate
energy of systems where the positively charged
particles have mass $M$ and the
negatively charged ones have mass $m$ to arbitrary
ratios

$m/M$. These bounds will prove binding, as long as

$(m/M)\leq 0.6$, including the case of  H$_2$.

The second proof is inspired by the Heitler-London
theory of binding in the
Born-Oppenheimer limit, corresponding to $m/M
\rightarrow 0$, and extends that theory
to an interval $0 \leq m/M \leq 0.144$.
For small values of $m/M$ it yields better bounds than
the first proof.

Details of these results, including a study of stability
as a function of the
masses of the constituent particles will be presented
in forthcoming papers by J.-M.R.\
\cite{Richard}  and M.S.\ \cite{Seifert}.

Throughout this paper, we shall employ units in which
the groundstate energy of
the actual hydrogen atom is $E_0 \left({ p  e^- }\right)=-
1$, and we restrict our analysis to Coulomb
systems, (except for a remark in the conclusions).

To determine the nature of binding in the H$_2$
molecule, one has first to understand
the ordering of the threshold energies that appear
when the system

$\left({ p p e^- e^- }\right)$ or $\left({ d p e^- e^-
}\right)$, is decomposed
into two or more non-interacting clusters. We only
consider the system

$\left({ p p e^- e^- }\right)$, but the arguments for a
system like
$\left({ d p e^- e^- }\right)$ are similar.

The minimal energy of a decomposition of $\left({ p p
e^- e^- }\right)$ into four
non-interacting clusters is obviously $E_4 = 0$, the
one of a decomposition into three
non-interacting clusters,

$\left({ p e^- }\right) (p) \left({  e^- }\right)$,
is $E_3 = E_0 \left({ p  e^- }\right)=-1$.

The third decomposition is into the clusters
$\left({  p e^- e^- }\right)$ and $ (p) $.

The minimal energy of the resulting system is

$E_2 = E_0 \left({  p e^- e^- }\right)$, where

$E_0 \left({  p e^- e^- }\right)$ is the groundstate
energy of the system
$ \left({  p e^- e^- }\right).$ Hill \cite{Hillp} has shown
that this system has exactly
one bound state with the electrons in a spin-singlet,
the groundstate of the H$^-$
ion. However, the groundstate energy $E_0 \left({  p e^-
e^- }\right)$ of the
H$^-$ ion is not known exactly. We shall need a
somewhat accurate lower bound on

$E_0 \left({  p e^- e^- }\right)$. The easiest such bound
is obtained by
neglecting the Coulomb interaction between the
electrons and the kinetic energy
of the proton. The lower bound is then twice the
groundstate energy of the
hydrogen atom where the reduced mass $\mu$ is
replaced by the electron mass $m
> \mu$, i.e., $E_0 \left({  p e^- e^- }\right) > - 2 -

{\mathop{\rm const. }\nolimits} 10^{-3}$, with

${\mathop{\rm const. }\nolimits} \approx

 {1/2}$. This bound is very bad and is insufficient for
our purposes. It
must be improved by taking into account the
electronic repulsion. This can be
done by using the projection method \cite{Thir}. By
$H_0$ we denote the Hamiltonian of the
system $ \left({  p e^- e^- }\right)$, but where the mass
$M$ of the positively
charged particle, the proton, is set to $\infty$ and {\it
without} the Coulomb
repulsion $V$, between the electrons. Let $H' = H_0 +
V$, and let $H^{(3)}$
denote the true Hamiltonian of the system $\left({  p e^-
e^- }\right)$.
Denoting the groundstate energy of a Hamiltonian $H$
by $E_0 (H)$, we then have
\begin{equation}\label{H3-1}
E_0 \left({ H^{(3)} }\right)> E_0

\left({ H' }\right) \geq E_0 \left({ H_0 }\right) +
\left\langle{ \psi_0, V^{- 1}  \psi_0 }\right\rangle^{- 1},

\end{equation}
where $\psi_0$ is the groundstate of $H_0 $.
Evaluating the right hand side of this
inequality yields the bound
\begin{equation}\label{H3-2}
E_0 \left({ H^{(3)} }\right) > - {5 \over 4}

\left({1 +  {m \over M} }\right),
\end{equation}
in our units. Thus
\begin{equation}\label{H3-3}
E_0 \left({ H^{(3)} }\right)  > - 2, \quad \mbox{for}\quad

{m \over M} \leq {3 \over 5}\ ,
\end{equation}
and
\begin{equation}\label{H3-4}
E_0 \left({ H^{(3)} }\right)  > - 1.2507
\end{equation}
if $m/M$
is given its physical value.

The fourth decomposition of the system
$ \left({p  p e^- e^- }\right)$ into non-interacting
clusters is

$\left({p  p e^-  }\right) \left({ e^- }\right)$. Its lowest
energy is given by

$E'_2 = E_0 \left({p  p e^- }\right)$, the groundstate
energy of the
hydrogen ion H$_2^{\ +}$.
Again, the stability of the H$_2^{\ +}$ ion has been
established by Hill\cite{Hillp}, but
the exact value of

$ E_0 \left({pp   e^- }\right)$ is not known.

Thus we must prove a suitable lower bound on

$ E_0 \left({pp   e^- }\right)$.
This can be accomplished by using the
Born-Oppenheimer approximation and the
so-called criterion of local energy\cite{Thir}.
Setting the proton mass $M$ to $\infty$ we obtain a
two-center problem, with the
protons separated by a distance $R$, for a single
quantum mechanical electron.
Let $H_R $ denote the Hamiltonian of this system. The
Perron-Frobenius theorem
guarantees that $H_R $ has a unique groundstate
$\varphi_R (\vec{\rm x})$ which is a
positive function of the electron position $\vec{\rm x}$.
The corresponding
groundstate energy is denoted by $E_R$. Let $\varphi
(\vec{\rm x})$ be a positive
function of $\vec{\rm x}$, and define
$E_R (\vec{\rm x}) : = \varphi (\vec{\rm x})^{- 1 }
\left({ H_R \varphi }\right) (\vec{\rm x})$. Then
\begin{equation}\label{Inf-x}
 \inf_{\vec{\rm x}}^{ }
E_R (\vec{\rm x}) \leq E_R \leq \

\sup_{ \vec{\rm x}}^{ }
E_R (\vec{\rm x}) ,
\end{equation}
as one easily shows. Thus
\begin{equation}\label{ppe-1}
 E_0 \left({pp   e^-}\right) >\
\min_{ R}^{ }E_R \geq

\ \min_{R}^{ }
\left(  \inf_{\vec{\rm x}}^{ }
E_R (\vec{\rm x})  \right)

\end{equation}

Choosing $\varphi (\vec{\rm x})$ to be
 proportional to  $\exp (- \alpha \mu) \cosh (\beta \nu)$,
with

$\mu =  (r_1 + r_2)/R $ and
$\nu =  (r_1 - r_2 )/R$,
 where $r_i$ is the distance between the electron
and the $i^{{\mathop{\rm th }\nolimits}}$ proton, one
obtains, after a tedious
calculation (using elliptic coordinates) and for an
optimal choice of  the
constants $\alpha$ and $\beta$, that
\begin{equation}\label{ppe-2}
E_0 \left({p p    e^- }\right) > - 2, \quad \mbox{for} \quad
{m \over M} \leq 0.605

\end{equation}
and
\begin{equation}\label{ppe-3}
E_0 \left({p p    e^- }\right) > - 1.623
\end{equation}
if  $m/M$ is given its physical value.
Details of the proof will appear in \cite{Seifert}.
Our bounds suggest that

$E_2 = E_0 \left({p e^-    e^- }\right) > E'_2 = E_0

 \left({p p    e^- }\right) $, and this can presumably be
proven by refining
the estimates  in \cite{Seifert}.

Finally, we consider the decompositon of the system

$ \left({p p  e^- e^- }\right) $ into two non-interacting
hydrogen atoms
$ \left({p   e^-  }\right)
\left({p   e^-  }\right)$. The lowest energy,

$E''_2$, of the resulting system is, of course,
\begin{equation}
\label{th-two-atoms}
E''_2 = 2 E_0 \left({p   e^-  }\right) = - 2

\end{equation}

We define
\begin{equation}
\label{Min-th}
E_c =\min \left\{ E_4, E_3, E_2, E'_2, E''_2 \right\}
\end{equation}
Since  $0 = E_4 > E_3 > E_2, E'_2 > E''_2 = - 2$, we
have that
$E_c = - 2$.The HVZ theorem \cite{ReedSimon} now
tells us that the continuous spectrum of the
Hamiltonian $H^{(4)}$ of the four-body system

$\left({p p e^-    e^- }\right)$ is the interval

$[E_c, \infty)$,  with $E_c = 2 E_0 \left({p   e^- }\right)$,

for $m /M\leq 0.6$. In particular, when

$m/M$ is given its physical  value then
\begin{equation}
\label{continuous}
\sigma_{
{\mathop{\rm continuous }\nolimits} }
\left({ H^{(4)} }\right) = [- 2, \infty).

\end{equation}
Thus, to prove that the system

$\left({p p e^-e^- }\right)$ has bound states, it suffices
to construct a
variational wave function, $\psi$, whith the property
that
\begin{equation}\label{def-stab}
\left\langle{\psi,   H^{(4)}  \psi }\right\rangle < - 2
.\end{equation}
We shall sketch two somewhat complementary
methods to construct a $\psi$ such that
(\ref{def-stab}) holds.

The first method to prove (\ref{def-stab}) starts from a
beautiful argument of Hylleraas and
Ore \cite{Ore} suggesting that the positronium
molecule

$\left({ e^+ e^+ e^- e^- }\right)$ is bound. We make use
of their results in a way
that has been outlined briefly in \cite{BadHonnef}.
They use the simple variational wave function

$\psi \left(\left(\lambda\vec{\rm x}_i\right)\right)$,
 where
\begin{eqnarray}
\label{Hyll-wf}
\psi\left(\left(\vec{\rm x}_i \right)\right)&=&
 \exp -{1 \over 2}

\left( r_{13} + r_{14} + r_{23} + r_{24}\right)\nonumber\\
& &\times\cosh

{\beta \over 2}

\left(r_{13} - r_{14} - r_{23} + r_{24}\right),
\end{eqnarray}
with $r_{ij} =

\left\vert\vec{\rm x}_i - \vec{\rm x}_j\right\vert$.
The scale parameter $\lambda$ is determined by the
virial theorem (which holds for
variational bounds \cite{Fock}), and the variational
energy $E_0 (\beta) $ can be calculated
explicitly \cite{Ore,Richard}.
As a result one finds that
\begin{equation}\label{Hyll-ener}
E_0

\left(e^+ e^+ e^- e^-\right) \leq

\min_{\beta}^{ }
E_0 (\beta) \simeq - 2.0168 E_0

\left(e^+  e^-\right),
\end{equation}
and the minimum is reached near $\beta^2 = 0.48.$
The threshold of the continuous spectrum in

$\left({ e^+ e^+ e^- e^- }\right) $ is expected to be twice
the groundstate energy
of positronium, i.e.,
\begin{equation}\label{posit-th}
E_c = 2 E_0 \left({ e^+  e^- }\right)

\end{equation}
but the equality between $E_c$ and

$2 E_0 \left({ e^+  e^- }\right) $ has not been rigorously
established for the
positronium molecule, yet. However, we neither need
this equality in the following
arguments, nor is it needed to prove binding for

$\left({ e^+ e^+ e^- e^- }\right)$.

Next, we rewrite the Hamiltonian $H^{(4)}$ of the
hydrogen molecule as
\begin{equation}\label{H-decomp}
H^{(4)} = H_S + H_A,

\end{equation}
where
\begin{eqnarray}
H_S &=&
\left({1 \over4M} + {1 \over4m}\right)
\left(\vec{\rm p}_1^2 + \vec{\rm p}_2^2 + \vec{\rm
p}_3^2 + \vec{\rm p}_4^2\right) + V \\
H_A&=&
\left({1 \over 4M} - {1 \over 4m}\right)
\left(\vec{\rm p}_1^2 + \vec{\rm p}_2^2 - \vec{\rm
p}_3^2 - \vec{\rm p}_4^2\right),

\end{eqnarray}
where $H_S$ is even under charge conjugation and
$H_A$ is odd; see \cite{BadHonnef}.
We now notice that the minimal energies in the spectra
of $H^{(4)}$ and of $H_S$
for two infinitely separated hydrogen atoms are
identical.
When $(m/M) \leq 0.6$ we know  that this is the lowest
threshold. It then follows immediately from
(\ref{Hyll-ener}) that the lowest energy in the spectrum
of $H_S$ satisfies

\begin{equation}
\label{Bound-HS}
E_0(H_S) \leq 2.0168 E_0\left(p e^- \right)

\end{equation}
which is strictly smaller than $-2$ if $m/M$ is given its
physical value.

Next, we note that
\begin{equation}
\label{Comp-H-HS}
E_0 \left( H^{(4)}\right)
 <  \left\langle \psi_0^S, (H_S+H_A)\psi_0^S
\right\rangle,

\end{equation}
where $\psi_0^S$ is a normalized wave function
symmetric under charge
conjugation, like the one used by Hylleraas and Ore,
and such that $\left\langle
\psi_0^S,H_S\psi_0^S \right\rangle <
E_0(H_S)+\epsilon$, for some $\epsilon > 0$ which
can be chosen arbitrarily small. Since $H_A$ is odd
under charge conjugation,

$\left\langle
\psi_0^S, H_A \psi_0^S \right\rangle = 0$. Hence it
follows that

\begin{equation}
\label{Stab1}
E_0

\left(H^{(4)}\right)
 <  E_0(H_S) \leq 2.0168 E_0\left(p e^- \right)

\end{equation}
which, by (\ref{def-stab}), proves that $(ppe^-e^-)$ is
bound, provided

$ (m/M) < 0.6$, in particular when

$ m/M$ is given its physical value.

The inequality $E_0\left(H^{(4)} \right) <E_0(H_S)$ is
reflected nicely in actual binding energies
which grow from 3 \% for the positronium molecule to
17 \% for the hydrogen molecule, according to the
estimates reported in \cite{Rebane}.

Previously, Abdel-Raouf \cite{Abdel} and Rebane
\cite{Rebane} stressed the regularity of the binding
energy as a
function of $ m/M$, but missed the fact that the bound
(\ref{Hyll-ener}) for $m = M$ implies binding
for $m/M$ small enough.

Let us finally sketch how the results on the thresholds
can be combined with an improved version of
Heitler-London theory to prove binding for the
hydrogen molecule. In nuclear center-of-mass
coordinates, the Hamiltonian for the hydrogen
molecule is given (in suitable units) by
\begin{eqnarray}
\label{def-H4}
 H^{(4)} &=& -{1\over M} \Delta_Y - {1\over 2 \mu}
\Delta_{y_1} - {1\over 2\mu} \Delta_{y_2} - {1\over
2M}\vec{\nabla}_{y_1} \cdot
\vec{\nabla}_{y_2}\nonumber\\
& &-2 \sum^2_{\scriptstyle i=1\atop\scriptstyle
\epsilon=\pm 1} \ {1\over | \vec{\rm y}_i +\epsilon
\vec{\rm Y}/2 |}
+{2\over |\vec{\rm Y}|} + {2\over |\vec{\rm y}_1 -
\vec{\rm y}_2 |} ,

\end{eqnarray}
with $\mu =
 2Mm /(2 M + m)$.
Within the Born-Oppenheimer approximation, suitable
ansatz wave functions

are the ones of Heitler and London given  by
\begin{eqnarray}
\label{Heitler-London}
\varphi_{\pm}(\vec{\rm Y}, \vec{\rm y}_1,\vec{\rm y}_2)
&=& C_\pm\left[
f\left(|\vec{\rm y}_1 -\vec{\rm Y}/2|\right)
f\left(|\vec{\rm y}_2 + \vec{\rm Y}/2|\right)
\right.\nonumber\\
& &\left. \pm f\left (|\vec{\rm y}_1 +  \vec{\rm Y}/2|\right
)
f\left(|\vec{\rm y}_2 - \vec{\rm Y}/2|\right )

\right]

\end{eqnarray}
with ``+'' corresponding to spin singlet and ``$-$''
corresponding to spin triplet for the electrons, and
$f(r) = \mbox{const}\times\exp(- 2 \mu r )$. The orbital
groundstate wave function of the hydrogen molecule
is symmetric in the electron positions. This
motivates us to use the following variational wave
function :
\begin{equation}
\label{Var-wf-2}
\psi\left (\vec{\rm Y},\vec{\rm y}_1,\vec{\rm y}_2\right )
= C(\beta) |\vec{\rm Y} | e^{-\beta |\vec{\rm Y}|}
\varphi_+ \left (\vec{\rm Y},\vec{\rm y}_1,\vec{\rm
y}_2\right),
\end{equation}
where $C(\beta)$ is a constant chosen such that
$\langle \psi,\psi \rangle = 1$.

After very tedious calculations and lengthy estimates
carried out in \cite{Seifert}, one finds that, for an
optimal choice of $\beta$,
\begin{equation}
\label{Stab-BH-type}
\langle \psi,H^{(4)} \psi \rangle < -2.082
\end{equation}
which proves (\ref{def-stab}) and hence establishes
the result that $(ppe^-e^-)$ is bound.

There is a variant of our methods which yields a very
simple proof of binding of
$\left({ p p e^- e^- }\right)$, for arbitrary values of

$  m / M$, but does not provide quantitative estimates
on binding energies: if the lowest threshold were
$\left(ppe^-\right)\left(e^-\right)$ or
$\left(pe^-e^-\right)\left(p\right)$, instead of
$\left(pe^-\right)\left(pe^-\right)$, the long-range
attraction would keep the ion and the charge together,
and the stability of  $\left({ p p e^- e^- }\right)$
can be proven.

 Our methods can be extended, in principle, to
systems of three and four particles interacting through
``universal'' (mass- and flavour-independent)
two-body potentials. Binding will be strongest in
systems of four particles with masses $(M,M,m,m)$ in
the limiting regime when $M \gg m$. Simple quark
models have flavour-independent potentials, and one
thus is led to predict stability of exotic
mesons $(QQ\bar q \bar q)$ with two units of heavy
flavour \cite{BadHonnef}.

J.-M.R.\ would like to thank A. Martin and T.T. Wu for
their continuous  interest, and F. Levin for a
fruitful discussion at Brown University. J.F., G.-M.G.\
and M.S.\  thank W. Hunziker for
useful remarks and encouragement.

J.F.\ and J.-M.R.\ thank

 the Institut des Hautes Etudes Scientifiques,
where this work was written, for hospitality.


\end {document}